\documentclass[lettersize,journal]{IEEEtran}
\usepackage{amsmath, amsthm, amssymb,amsfonts}
\usepackage{algorithmic}
\usepackage{algorithm}
\usepackage{array}
\usepackage{textcomp}
\usepackage{stfloats}
\usepackage{url}
\usepackage{verbatim}
\usepackage{graphicx}
\usepackage{caption}
\usepackage{subcaption}
\usepackage{cite}
\usepackage[usenames,dvipsnames]{color}
\begin{document}

\title{On the Spatial-Wideband Effects in Millimeter-Wave Cell-Free Massive MIMO}

\author{
    Seyoung Ahn,~\IEEEmembership{Student Member,~IEEE},~Soohyeong Kim,~Yongseok Kwon,~Joohan Park,~Jiseung Youn,~and~Sunghyun Cho,~\IEEEmembership{Member,~IEEE}
    \thanks{Manuscript received MM DD, YYYY; revised MM DD, YYYY.}
    \thanks{S. Ahn, S. Kim, Y. Kwon, and J. Youn are with the Department of Computer Science and Engineering, Major in Bio Artificial Intelligence, Hanyang University, Ansan, South Korea (e-mail: tpdud1014@hanyang.ac.kr; dreammusic23@hanyang.ac.kr, totoey200@hanyang.ac.kr, yjs1104@hanyang.ac.kr).}%
    \thanks{J. Park and S. Cho are with the Department of Computer Science and Engineering, Hanyang University, Ansan, South Korea (e-mail: 1994pjh@hanyang.ac.kr; chopro@hanyang.ac.kr).}
}

\markboth{Journal of \LaTeX\ Class Files,~Vol.~14, No.~8, August~2023}%
{Shell \MakeLowercase{\textit{et al.}}: A Sample Article Using IEEEtran.cls for IEEE Journals}

\IEEEpubid{0000--0000/00\$00.00~\copyright~YYYY IEEE}

\maketitle

\begin{abstract}
In this paper, we investigate the spatial-wideband effects in cell-free massive MIMO (CF-mMIMO) systems in mmWave bands. The utilization of mmWave frequencies brings challenges such as signal attenuation and the need for denser networks like ultra-dense networks (UDN) to maintain communication performance. CF-mMIMO is introduced as a solution, where distributed access points (APs) transmit signals to a central processing unit (CPU) for joint processing. CF-mMIMO offers advantages in reducing non-line-of-sight (NLOS) conditions and overcoming signal blockage. We investigate the synchronization problem in CF-mMIMO due to time delays between APs. It proposes a minimum cyclic prefix length to mitigate inter-symbol interference (ISI) in OFDM systems. Furthermore, the spatial correlations of channel responses are analyzed in the frequency-phase domain. The impact of these correlations on system performance is examined. The findings contribute to improving the performance of CF-mMIMO systems and enhancing the effective utilization of mmWave communication.
\end{abstract}


\section{Introduction}
\IEEEPARstart{A}{s} mobile communication systems evolved, and the demands for communication technologies utilizing extremely high-frequency bands such as millimeter-wave (mmWave) and terahertz (THz) bands have been aroused. For the utilization of the higher-frequency band, we should consider that the signal can be easily attenuated by the obstacle or distance between transceivers. Attenuating signals by distance may decrease cell coverage, and communication systems require densified networks, namely, ultra-dense networks (UDN), to maintain communication performance. However, in the UDN, some potential drawbacks of the multi-cell networks may be deepened, such as inter-cell interference (ICI) or pilot contamination. 

Several cooperative communication methods, such as the coordinated multi-point (CoMP) and network MIMO system, have been studied to mitigate ICI and improve communication performance in the cell-edge region. However, cooperative communications suffer from synchronizing all serving BSs because the serving BSs share the channel state information and independently process the received signals by their own RF chain. Cell-free massive MIMO (CF-mMIMO) has been proposed as the complete form of cooperative communication to address the synchronization problem. Specifically, in CF-mMIMO, a single central processing unit (CPU) processes the transceived signals from the users by the distributed BSs called the access points (APs) connected by the fronthaul links. The distributed structure of CF-mMIMO yields another advantage for the higher-frequency band communications to reduce the non-line-of-sight (NLOS) probability. As the frequency band increases, the diffraction becomes more severe due to the shorter wavelength, and even signal blockage may occur. The CF-mMIMO system places APs in closer proximity to the user compared to the conventional massive MIMO BS in cellular networks. It can decrease the distance between AP and the user and consequently provides the same effect as removing the obstacles.

Most existing studies on CF-mMIMO systems tend to assume perfect synchronization, but in higher frequency bands, the synchronization issues may still remain. In massive MIMO systems of cellular networks, the signal processing only considers the phase difference because distances between UE and antennas are almost the same. Adversely, CF-mMIMO is a form in which numerous antennas in the antenna array of cellular massive MIMO are distributed as the APs throughout the system. To consider this distributed nature of CF-mMIMO, the signal processing for CF-mMIMO systems should consider phase differences caused not only by the antenna in each AP but also by the different distances between UE and APs.



In this letter, we first investigate the synchronization problem due to the time delays between different APs. The time delays are not negligible and may cause severe inter-symbol interference (ISI) in the OFDM system. We provide the minimum length of the cyclic prefix for reducing the ISI. Furthermore, we investigate the spatial correlations for the channel responses in the frequency-phase domain. We decompose the spatial correlation matrix into two types of correlations, such as the \emph{micro-correlation} in the antenna of each AP and the \emph{macro-correlation} among APs. Consequently, we investigate the effect of correlations by introducing the numerical results.

\IEEEpubidadjcol

\textbf{Notation}: Boldface lowercase and uppercase letters, such as $\textbf{x}$ and $\textbf{X}$, denote column vectors and matrices, respectively. The superscripts $\textrm{T}$, $*$, and $\textrm{H}$ denote the transpose, conjugate, and conjugate transpose, respectively. The matrix $\textrm{diag}(\textrm{x})$ denotes the diagonal matrix with the elements in the vector $\textrm{x}$ on the diagonal. The $M$-dimensional identity matrix and the vector with all-one elements denote $\mathbf{I}_{M}$ and $\mathbf{1}_{M}$, respectively. We use $\triangleq$ for definitions. The complex set with the $M \times N$ dimensional elements denotes $\mathbb{C}^{M \times N}$. The function $\textrm{Re}\{\cdot\}$ stands for the real part of a complex signal. The operators $\otimes$ and $\circ$ stand for the Kronecker and Hadamard products, respectively. The expectation of the matrix $\textrm{X}$ denotes $\mathbb{E}\{\textrm{X}\}$.

\section{System model}
\begin{figure}[t]
    \centering
    \includegraphics[width=\columnwidth]{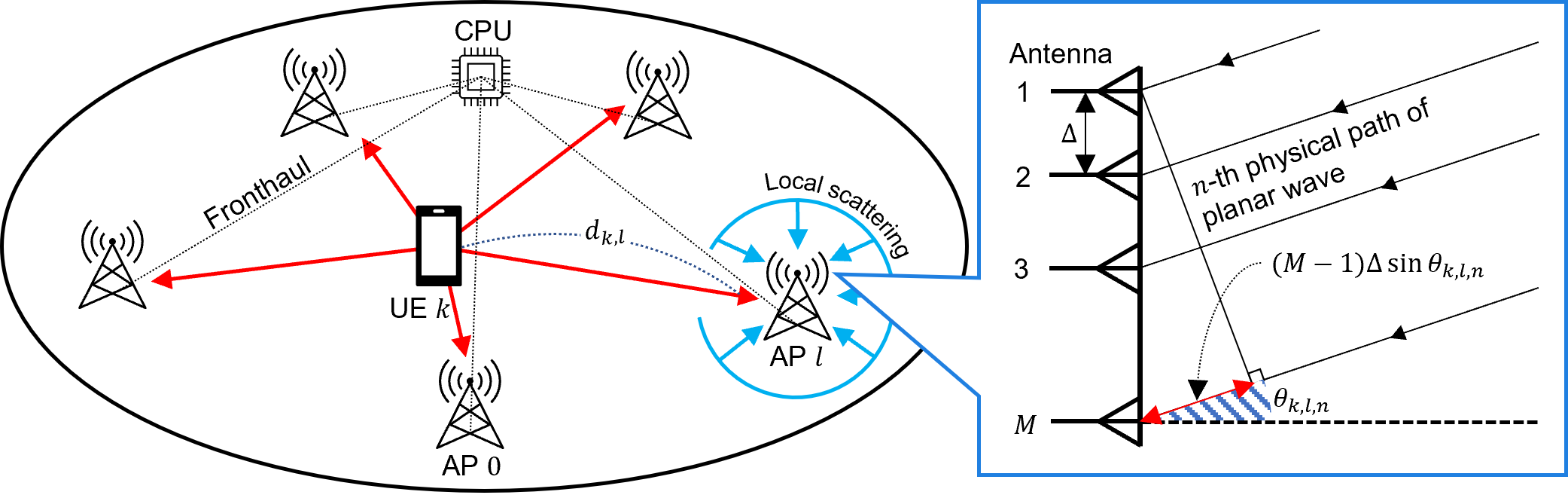}
    \caption{System model.}
    \label{fig:system_model}
\end{figure}

Considering a CF-mMIMO system consisting of $L$ APs and $K$ UEs equipped with a single antenna as illustrated in the left-hand side of Figure \ref{fig:system_model}. Each AP is equipped with a $M$-antenna uniform linear array (ULA) and connected to a CPU via a fronthaul link. Each antenna is separated by an antenna spacing $\triangle$. We assume that the CF-mMIMO system works in a time division duplex (TDD) protocol. In a TDD protocol, we only consider the uplink channel training because the uplink channel state information (CSI) can be utilized for both the uplink and downlink transmission by channel reciprocity. Orthogonal frequency-division multiplexing (OFDM) with $P$ subcarriers is adopted. In the OFDM system, the carrier frequency and wavelength denote $f_{c}$ and $\lambda_{c}$, respectively. The subcarrier spacing will be $\eta=W/P$ when the transmission bandwidth is $W$.
We consider that UE $k$ sends the planar wave to the AP $l$ with the $N$ physical paths. The direction of arrival (DoA) of the signal from UE $k$ to the AP $l$ is denoted by $\theta_{k, l, n}$ for the $n$-th physical path. The schematic descriptions of the signal model in AP $l$ are illustrated by the right-hand side of Figure \ref{fig:system_model}.\par

First of all, we assume that the first antenna of the nearest AP to UE $k$ is perfectly synchronized, and other antennas and APs are received the delayed version of the signal. we define the time delay $\tau_{k, l, m, n}$ of the $m$th antenna of the AP $l$. Specifically, $\tau_{k, l, m, n}$ consists of two types of delay caused by the distance $d_{k, l}$ between user $k$ and AP $l$ and the adjacent antennas at AP $l$ that can be explicitly computed as $\frac{d_{k, l}}{c}$ and $m\frac{\triangle\sin{\theta_{k, l, n}}}{c}$, respectively. Note that $c$ denotes the speed of light. Therefore, we can represent the time delay as follows:
\begin{equation}
\label{eq:time_delay}
    \tau_{k, l, m, n}=\frac{d_{k, l}}{c} + m\frac{\triangle\sin{\theta_{k, l, n}}}{c},
\end{equation}

The baseband signal generated by the user $k$ can be represented as
\begin{equation}
\label{eq:transmitted_baseband_signal}
    x_{k}(t) = \sum_{i=-\infty}^{+\infty}s_{k}[i]\delta(t-iT_{s}),
\end{equation}
where $\delta(\cdot)$ is the pulse shaping function. The user transmits this signal by modulating to the corresponding passband signal as $\textrm{Re}\{x_{k}(t)e^{j2 \pi f_{c}t}\}$. The antenna $m$ at the AP $l$ receives the delayed passband signal in the $n$th path as
\begin{equation}
\label{eq:received_passband_signal}
    \begin{aligned}
        & \textrm{Re}\{\Bar{\alpha}_{k, l, n}x_{k}(t - \tau_{k, l, m, n})e^{j2 \pi f_{c}(t - \tau_{k, l, m, n})}\} \\
        & = \textrm{Re}\{\Bar{\alpha}_{k, l, n}x_{k}(t - \tau_{k, l, m, n})e^{j2 \pi f_{c}(t - \frac{d_{k, l}}{c} - m\frac{\triangle\sin{\theta_{k, l, n}}}{c}})\} \\
        & = \textrm{Re}\{\Bar{\alpha}_{k, l, n}x_{k}(t - \tau_{k, l, m, n})e^{-j2 \pi \frac{d_{k, l}}{\lambda_{c}}} e^{-j2 \pi m\frac{\triangle\sin{\theta_{k, l, n}}}{\lambda_{c}}} e^{j2 \pi f_{c}t}\},
    \end{aligned}
\end{equation}
where $\Bar{\alpha}_{k, l, n}$ is the corresponding complex channel gain.
Then, we can formulate the corresponding received baseband signal at the $t$ time instant as
\begin{equation}
\label{eq:received_baseband_signal}
    y_{k, l, m}(t) = \sum_{n=1}^{N}\Bar{\alpha}_{k, l, n}x_{k}(t - \tau_{k, l, m, n})e^{-j2 \pi \frac{d_{k, l}}{\lambda_{c}}} e^{-j2 \pi m\frac{\triangle\sin{\theta_{k, l, n}}}{\lambda_{c}}}.
\end{equation}

\section{Spatial-wideband effects of mmWave cell-free massive MIMO systems}
In this section, we investigate the spatial-wideband effect of cell-free massive MIMO systems utilizing mmWave bands. Specifically, we formulate the channel model with the spatial-wideband effects. Based on the channel model, we analyze the spatial-wideband effects in three different aspects: delay spread in the angular-delay domain, frequency spread in the frequency-phase domain, and spatial correlations. In the analysis, we confirm the synchronization problem and errors in the frequency channel responses.\par

As discussed previously, if the transmission bandwidth and antenna spacing are narrow or the number of antennas is small, the ISI can be negligible due to the sufficiently short time delay compared to the symbol period. However, the cell-free massive MIMO system has a similar structure where the existing massive MIMO antennas are spread throughout the network in the form of APs, resulting in a large number of antennas and very wide spacing between them. Furthermore, the mmWave band can be worked as an obstacle by amplifying the effect of spatial-wideband effects although the cell-free massive MIMO is one of the key technologies to enable the high-frequency bands.

\subsection{Channel model}
The uplink spatial-time channel of user $k$ at the $m$th antenna of AP $l$ can be modeled as
\begin{equation}
\label{eq:spatial-time_channel}
    \begin{aligned}
        & \left[\mathbf{h}^{\textrm{ST}}_{k, l}(t)\right]_{m}\\
        & = \sum_{n=1}^{N}\Bar{\alpha}_{k, l, n}\delta(t - \tau_{k, l, m, n})e^{-j2 \pi \frac{d_{k, l}}{\lambda_{c}}}e^{-j2 \pi m\frac{\triangle\sin{\theta_{k, l, n}}}{\lambda_{c}}} \\
        & = \sum_{n=1}^{N}\alpha_{k, l, n}\delta(t - \tau_{k, l, m, n})e^{-j2 \pi m\frac{\triangle\sin{\theta_{k, l, n}}}{\lambda_{c}}},
    \end{aligned}
\end{equation}
where the corresponding complex channel gain is $\alpha_{k, l, n} \triangleq \Bar{\alpha}_{k, l, n}e^{-j2 \pi \frac{d_{k, l}}{\lambda_{c}}}$. Then, based on the continuous-time Fourier transform, we can obtain the spatial-frequency channel response as
\begin{equation}
\label{eq:spatial-frequency_channel}
    \begin{aligned}
        & \left[\mathbf{h}^{\textrm{SF}}_{k, l}(f)\right]_{m} = \int_{-\infty}^{+\infty}\left[\mathbf{h}^{\textrm{ST}}_{k, l}(t)\right]_{m}e^{-j2 \pi ft}dt \\
        & = \sum_{n=1}^{N}\alpha_{k, l, n}e^{-j2 \pi m\frac{\triangle\sin{\theta_{k, l, n}}}{\lambda_{c}}}e^{-j2 \pi f\frac{d_{k, l}}{{c}}}e^{-j2 \pi fm\frac{\triangle\sin{\theta_{k, l, n}}}{c}}
    \end{aligned}
\end{equation}
By the obtained response and \cite{ref:benchmark}, we can confirm the extra component of the phase shift represented as $e^{-j2 \pi f\frac{d_{k, l}}{{c}}}$. Different from the conventional massive MIMO system, we can get an insight that implementing the cell-free massive MIMO systems with OFDM should not neglect the synchronization problem due to the time delay among APs. Based on the \eqref{eq:spatial-frequency_channel}, we can get the total channel response of the user $k$ for the $L$ APs in the entire systems as
\begin{equation}
    \mathbf{h}_{k}(f) = \textrm{vec}(\left[\mathbf{h}^{\textrm{SF}}_{k, 0}(f), \mathbf{h}^{\textrm{SF}}_{k, 1}(f), ..., \mathbf{h}^{\textrm{SF}}_{k, L-1}(f)\right]^{\textrm{T}}) \in \mathbb{C}^{LM \times 1}.
\end{equation}
We formulate two vectors such as the complex gain vector $\boldsymbol{\alpha}_{k, n} \in \mathbb{C}^{L \times 1}$ and macro-steering vector $\mathbf{d}_{k}(f) \in \mathbb{C}^{L \times 1}$ respectively as
\begin{equation}
\label{eq:complex_gain_vec}
    \boldsymbol{\alpha}_{k, n} = \left[\alpha_{k, 0, n}, \alpha_{k, 1, n}, ..., \alpha_{k, L-1, n}\right]^{\textrm{T}},
\end{equation}
\begin{equation}
\label{eq:macro_steering_vec}
    \mathbf{d}_{k}(f) = \left[e^{-j2 \pi f\frac{d_{k, 0}}{c}}, e^{-j2 \pi f\frac{d_{k, 1}}{c}}, ..., e^{-j2 \pi f\frac{d_{k, L-1}}{c}}\right]^{\textrm{T}}.
\end{equation}
Moreover, $\Theta(f)$ stands for the $(L \times M)$-dimensional phase-shift matrix whose $(l, m)$-th element is
\begin{equation}
\label{eq:phase-shift_matrix}
    \left[\Theta_{n}(f)\right]_{l, m} \triangleq e^{-j2 \pi f_{c}\left(1 + \frac{f}{f_{c}}\right)\frac{m\triangle\sin\theta_{k, l, n}}{c}}.
\end{equation}
Based on \eqref{eq:complex_gain_vec}, \eqref{eq:macro_steering_vec}, and \eqref{eq:phase-shift_matrix}, we can represent the spatial-channel response in \eqref{eq:spatial-frequency_channel} as follows:
\begin{equation}
\label{eq:arranged_channel_responses}
    \mathbf{h}_{k}(f) = \sum_{n=1}^{N}\textrm{vec}(\textrm{diag}(\boldsymbol{\alpha}_{k, n}\circ\mathbf{d}_{k}(f))\Theta_{n}(f)).
\end{equation}
In \eqref{eq:arranged_channel_responses}, the macro-steering vector and phase-shift matrix are frequency-dependent, which stands for the beam-squint effect.\par

\subsection{Beam-squint effects}
We then investigate the spatial-wideband effects for the macro-steering vector and phase-shift matrix by transforming \eqref{eq:macro_steering_vec} and \eqref{eq:phase-shift_matrix} to the virtual angle domain by discrete Fourier transform (DFT)\cite{ref:beam_squint, ref:spatial_wideband_effect}. Let $F_{L}$ be the $L$-dimensional normalized DFT matrix.

\subsection{Inter-symbol interference in OFDM system}
The difference in the time delay may cause the synchronization error because a single data symbol can arrive at each AP at different time instances in the CF-mMIMO system. The data symbols arrived at different times may cause inter-symbol interference without sufficient cyclic prefix (CP) length. In this section, we introduce the minimum CP length to remove the inter-symbol interference. Let $\psi_{k, l}$ denote the phase shift for the frequency $f$ in the spatial-frequency channel responses as follows
\begin{equation}
\label{eq:phase_shift}
    \psi_{k, l} = e^{-j2 \pi f\left(\frac{d_{k, l}}{c}+\frac{m\triangle\sin{\theta_{k, l, n}}}{c}\right)}.
\end{equation}
The delay difference for all subcarriers can denote as
\begin{equation}
\label{eq:delay_difference}
    \begin{aligned}
        \tau_{k, l, m}^{P} & = P\eta\left(\frac{d_{k, l}}{c}+\frac{m\triangle\sin{\theta_{k, l, n}}}{c}\right)\\
        & = W\left(\frac{d_{k, l}}{c}+\frac{m\triangle\sin{\theta_{k, l, n}}}{c}\right).
    \end{aligned}
\end{equation}
Assuming that indices of APs are sorted by the distances from the UE and the following inequations hold
\begin{equation}
    \label{eq:distances}
    d_{k, 0} \leq d_{k, 1} \leq ... \leq d_{k, L-1}.
\end{equation}
The minimum CP length should be larger than the difference between the maximum and minimum delay differences. Therefore, we can calculate the minimum CP length $\textrm{CP}_{\textrm{min}}$ as follows
\begin{equation}
    \begin{aligned}
        \textrm{CP}_{\textrm{min}} & \geq \left|\tau_{k, L-1, M-1}^{P} - \tau_{k, 0, 0}^{P}\right|\\
        & = W\left|\frac{d_{k, L-1}-d_{k, 0}}{c}+\frac{(M-1)\triangle\sin{\theta_{k, L-1, n}}}{c}\right|\\
        & \approx W\left|\frac{d_{k, L-1}-d_{k, 0}}{c}\right|.
    \end{aligned}
\end{equation}
Here, the delay difference among the antenna elements is ignorable compared with the difference among the APs.
\section{Numerical results}
We employ the UMi-Street Canyon NLOS channel model introduced in 3GPP TR 38.901 \cite{ref:channel_model}.

\section{Conclusion}

\bibliographystyle{IEEEtran}
\bibliography{main}

\vfill

\end{document}